%% file: Kaisin_n_en.tex
\documentclass[
aps,%
11pt,%
final,%
notitlepage,%
oneside,%
onecolumn,%
nobibnotes,%
nofootinbib,%
superscriptaddress,%
noshowpacs,%
centertags]%
{revtex4}
\input{sao_cmd_author.tex}

\begin{document}
\selectlanguage{english}





\title{Star Formation in Nearby Dwarf Galaxies\footnote{Based on observations with the
6-m telescope of the Special Astrophysical Observatory of the
Russian Academy of Sciences.}}

\author{\firstname{S.~S.}~\surname{Kaisin}}
\email{skai@sao.ru}
\affiliation{\saoname}
\author{\firstname{I.~D.}~\surname{Karachentsev}}

\received{July 20, 2018}%
\revised{December 12, 2018}%
\accepted{December 12, 2018}%

\begin{abstract}
We report the measured H$\alpha$ fluxes and images of 66 nearby
objects observed with the 6\mbox{-}m telescope of the Special
Astrophysical Observatory of the Russian Academy of Sciences.
Three of these objects---IC\,2233, UGC\,4704, and NGC\,3432---are
late-type spiral galaxies, six objects are distant globular
clusters of the M31 galaxy, and the remaining ones are dwarf
galaxies. We used the measured H$\alpha$ fluxes to estimate the
integrated and specific star-formation rates and analyzed some of
the main features of star formation in dwarf galaxies and
late-type spirals based on a sample of more than 500 Local-volume
galaxies.
\end{abstract}
\maketitle


\section{INTRODUCTION}
Over the last decade observations of nearby galaxies aimed at
determining their star-formation rate (SFR) from H$\alpha$
emission flux have been carried out at the Special Astrophysical
Observatory of the Russian Academy of Sciences. The H$\alpha$
survey targets are Local-Volume  (LV) galaxies located within
11~Mpc. The results of our previous observations are reported in a
series of~9 papers references to which can be found in the last
paper of this series~\cite{kar2015a:Kaisin_n}.

Mass sky surveys made at optical wavelengths and in the H\,I 21~cm
line gave rise to a rapid increase of the number of galaxies in
the LV. Their number in the ''Updated Nearby Galaxy Catalog''
(UNGC)~\cite{kar2013a:Kaisin_n} reached 869, and more than 80\% of
the sample are made up of dwarf galaxies with stellar masses $\log
(M^*/M_{\odot}) < 9.0$. As of mid-2018 the LV galaxy
database~\cite{kai2012:Kaisin_n} included 1153
objects\footnote{Various data about these objects are available
at~\url{http://www.sao.ru/lv/lvgdb}}. H$\alpha$-images for a total
of more than 300 nearby galaxies were acquired with the 6-m
telescope of the Special Astrophysical Observatory of the Russian
Academy of Sciences within the framework of our  H$\alpha$-survey
program. Our survey, combined with other similar surveys
\cite{jam2004:Kaisin_n,hun2004:Kaisin_n,ken2008:Kaisin_n}, makes
up a more than 500-object sample of LV galaxies with measured
H$\alpha$. Most of these galaxies have their ultraviolet fluxes
measured with GALEX~\cite{gil2007:Kaisin_n} space telescope,
making it possible to estimate the star-formation rates in
galaxies on the time scale of about  100~Myr, which is about one
order of magnitude longer than the time interval gauged by
star-formation rate estimates based on H$\alpha$ emission. A
comparison of the two star-formation rates, SFR$_{\rm H\alpha}$
and SFR$_{\rm FUV}$, makes it possible to reveal starburst and
star-formation rate deceleration on~10--100~Myr long time
intervals.

Below we present the H$\alpha$-images and the integrated H$\alpha$
fluxes for 66 LV objects. Almost all these objects are dwarf
galaxies except for three late-type  spirals: IC\,2233, UGC\,4704,
and NGC\,3432, and six globular clusters at the distant (farther
than 100~kpc) periphery of the Andromeda galaxy (M31). Only six
galaxies of this list had their H$\alpha$-fluxes measured earlier,
and in three cases (JKB\,83, Grapes and NGC\,6503-d1) we found
greater than one order of magnitude errors in previous
H$\alpha$-flux measurements.

\begin{figure} 
\setcaptionmargin{5mm} \onelinecaptionsfalse \captionstyle{normal}
\includegraphics[scale=0.75]{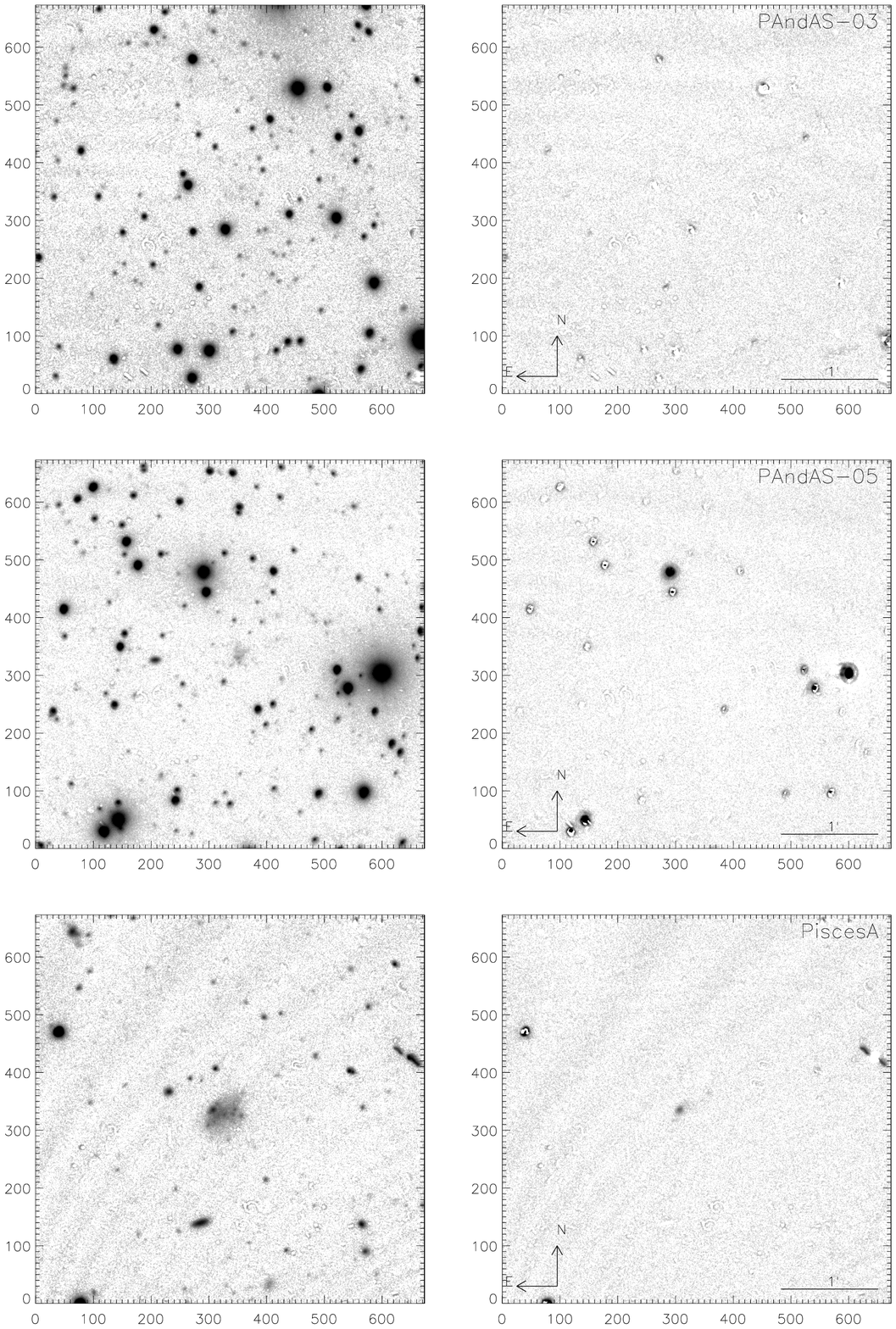}
\caption{Atlas of LV object images. The left-side images in each
pair are coadded H$\alpha$ and continuum exposures, and the
right-side images show the differences of these exposures. The
angular scale and orientation are indicated in the right-hand
images. Here we show only the first page of the image. The
complete image files are available at~{\tt
https://www.sao.ru/hq/leac/recent\_results/Halpha\_atlas\_66.pdf}.}
\label{fig1:Kaisin_n}
\end{figure}
\begin{table}
 \setcaptionmargin{0mm} \onelinecaptionstrue \captionstyle{normal}
\caption{General parameters of observed galaxies} 
\begin{tabular}{l|c|c|r|r|r|r|c|r}
\hline
Name      &  RA (2000) Dec & $B_T$,&   T &  $D$, &  $\log F_{H\alpha} $& $\log SFR_{H\alpha}$,& $\log SFR_{FUV}$,& $\log sSFR_{H\alpha}$,\\ 
		&	      & mag   & &  Mpc   &      &   [$M_{\odot}$\,yr$^{-1}$]&    [$M_{\odot}$\,yr$^{-1}$]&     [yr$^{-1}$]\\
		
\hline
    (1)          &    (2)    &  (3)  & (4)& (5)  &   (6)  &    (7)  &     (8)  &       (9)\\
\hline
PAndAS$-$03  & 000356.4+405319 &21.3  &$-$3   &0.78  &$<-$15.10  &$<-$6.26    &$<-$6.35     &$<-$10.64\\
PAndAS$-$05  & 000524.1+435535 &20.3  &$-$3  & 0.78  &$<-$15.30  &$<-$6.46    &$<-$6.38     &$<-$11.21\\
PiscesA    &   001446.0+104847 &18.5  &10  & 5.65 &$-$14.25 &$-$3.67      &$-$     &$-$10.20\\
JKB129     &   002041.4+083701 &17.4  &10  & 7.50 &$-$13.87 &$-$3.03   &$-$2.69    &$-$10.27\\
PAndAS$-$50  & 010150.6+481819 &19.3  &$-$3 & 0.78  &$<-$15.12  &$<-$6.21    &$<-$6.10     &$<-$11.49\\
PiscesB    &   011911.7+110718 &17.8  &10   &8.91   &$-$13.94   &$-$3.02     &$-$2.80      &$-$10.13\\
PAndAS$-$55  & 011920.4+460312 &20.0  &$-$3  & 0.78  &$<-$15.16  &$<-$6.33    &$<-$6.40     &$<-$11.20\\
AGC112454  &   013259.7+142224 &17.5   &9  &10.20 &  $-$13.39 &$-$2.35     &$-$      &$-$ 9.70\\
AGC114027  &   013441.8+143840 &18.8  &10  & 9.90 &$-$13.87 &$-$2.85     &$-$      &$-$ 9.66\\
AGC112503  &   013800.3+145858 &18.0  & 9  &10.20 &$-$13.61 &$-$2.57     &$-$      &$-$ 9.72\\
N672dwB    &   014711.1+274100 &21.0  &10   &7.20  &$<-$15.24  &$<-$4.48    &$<-$4.46     &$<-$10.17\\
N672dwA    &   014719.1+271516 &19.8  &10  & 7.20  &$<-$15.21  &$<-$4.44     &$-$4.48     &$<-$10.62\\
Triangulum$-$II& 021317.4+361042 &16.5  &$-$2 & 0.03  &$<-$15.17  &$<-$9.16       &$-$       &$<-$12.61\\
AGC123352  &   024839.2+231626 &19.2  &10   &8.27 &  $-$14.01 &  $-$3.00   &  $-$3.28   &   $-$ 9.77\\
AGC124056  &   025736.4+234721 &19.0  &10   &7.37  &$<-$15.20  &$<-$4.36    &$<-$4.23     &$<-$10.98\\
N1156dw1   &   030018.2+251456 &19.6  &10   &7.80  &$<-$15.11  &$<-$4.10    &$<-$3.76     &$<-$10.75\\
N1156dw2   &   030028.0+251817 &20.0  &10   &7.80  &$<-$15.25  &$<-$4.23   &  $-$3.21     &$<-$10.74\\
GALFA$-$Dw4  & 054544.8+104616 &17.8  &10   &7.22 &  $-$13.24  & $-$2.05    &   $-$      &  $-$ 9.81\\
MADCASH    &   074238.9+652502 &20.5  &$-$2   &3.39  &$<-$15.26  &$<-$5.19    &$<-$5.26     &$<-$11.05\\
IC2233     &   081358.9+454432 &13.1  & 7  &12.20  &$-$12.11  &$-$0.75    &$-$0.26     &$-$ 9.97\\
LV J0831+4104& 083141.2+410454 &17.7  & 9  &7.90  &$-$15.25  &$-$4.44    &$-$3.26     &$-$11.46\\
LV J0843+4025& 084338.0+402547 &17.9  & 9  &7.80  &$-$13.89  &$-$3.10    &$-$2.99     &$-$10.02\\
N2683dw2   &   085420.5+331458 &19.6 &$-$2 & 9.82  &$<-$15.28  &$<-$4.29    &$<-$4.34     &$<-$11.43\\
UGC04704    &  085900.3+391236 &15.0  & 8 &15.20  &$-$12.84  &$-$1.38    &$-$0.89     &$-$10.31\\
AGC198507  &   091525.8+252510 &18.6  & 9 & 8.90  &$-$15.01: &$-$4.10:   &$-$3.20     &$-$10.87:\\
JKB83      &   095549.6+691957 &19.7  &10  &3.70  &$-$13.67  &$-$3.49    &$-$4.24     &$-$ 9.14\\
PGC30114   &   101843.0+460244 &15.7  & 8  &8.05  &$-$12.54  &$-$1.74    &$-$1.88     &$-$ 9.63\\
UGC05571   &   101942.4+520356 &16.5  & 9 & 8.24  &$-$13.39  &$-$2.57    &$-$2.24     &$-$10.06\\
LV J1028+4240& 102833.0+424008& 17.4  &10 & 7.55  &$-$14.13  &$-$3.38    &$-$2.67     &$-$10.44\\
PGC2277751  &  103512.1+461412& 17.4  & 9  &7.25  &$-$14.57  &$-$3.85    &$-$3.15     &$-$10.90\\
N3344dw1   &   104244.0+250130 &20.0  &10  &9.82  &$<-$15.30  &$<-$4.31    &$<-$4.33     &$<-$10.60\\
Willman1   &   104921.0+510300 &16.0 &$-$2  &0.04  &$<-$15.22  &$<-$9.03    &$-$8.51     &$<-$12.75\\
NGC3432    &   105231.1+363708 &11.7  & 8 & 9.20  &$-$11.24  &$-$0.15    &$-$0.00     &$-$ 9.79\\
LV J1052+3639& 105240.8+363954 &15.8  &10 & 9.20  &$-$12.86  &$-$1.94    &$-$2.01     &$-$ 9.82\\
HS1053+3624 &  105640.4+360828 &17.8  &10 & 9.20  &$-$13.51  &$-$2.58    &$-$2.54     &$-$ 9.69\\
VV747      &   105747.0+361539 &15.5  &10 & 9.20  &$-$12.63  &$-$1.70    &$-$1.85     &$-$ 9.72\\
PGC034671  &   111948.6+554322 &16.5  & 9 & 9.90  &$-$14.22  &$-$3.24    &$-$2.81     &$-$10.91\\
UGC06757   &   114659.1+612006 &16.3  &10 & 4.61  &$-$14.12  &$-$3.78    &$-$3.02     &$-$10.90\\
\hline
\end{tabular}
\end{table}

\setcounter{table}{0}
 \begin{table}
\setcaptionmargin{0mm} \onelinecaptionstrue \captionstyle{normal}
 \caption{}
 \begin{tabular}{l|c|c|r|r|r|r|c|r}
\hline
Name      &  RA (2000) Dec & $B_T$,&   T &  $D$, &  $\log F_{H\alpha}$& $\log SFR_{H\alpha}$,& $\log[SFR_{FUV}$,& $\log sSFR_{H\alpha}$\\ 
		&	      & mag   & &  Mpc   &      &   [$M_{\odot}$\,yr$^{-1}$]&    [$M_{\odot}$\,yr$^{-1}$]&     [yr$^{-1}$]\\
\hline
    (1)          &    (2)    &  (3)  & (4)& (5)  &   (6)  &    (7)  &     (8)  &       (9)\\
\hline

Grapes     &   115205.6+544732 &18.5  &10 & 5.58  &$-$14.63  &$-$4.15      &$-$       &$-$10.51\\
PGC3401153 &   115352.4+512938 &17.5  & 9 & 7.79  &$-$14.63  &$-$3.85    &$-$3.06     &$-$10.90\\
MAPS1206+31&   120634.5+312033 &15.8  & 8 & 7.43  &$-$13.78  &$-$3.04    &$-$2.97     &$-$10.85\\
UGC07320   &   121728.5+444841 &15.4  &10 & 9.20  &$-$13.87  &$-$2.95    &$-$2.61     &$-$10.99\\
KK135      &   121934.7+580234 &18.1  &10 & 4.46  &$-$13.84  &$-$3.54    &$-$3.36     &$-$9.86\\
PGC5059199 &   122409.9+261352 &19.8  &10 & 3.37  &$-$14.59  &$-$4.53      &$-$       &$-$9.97\\
AGC724906  &   123056.0+263040 &17.4  &10 & 7.55  &$-$15.17  &$-$4.41    &$-$3.18     &$-$11.50\\
DDO133     &   123253.0+313221 &13.5  &10 & 4.88  &$-$12.36  &$-$1.98    &$-$1.73     &$-$10.22\\
PGC041749  &   123352.7+393733 &16.5  & 9 & 8.24  &$-$13.46  &$-$2.64    &$-$2.11     &$-$10.15\\
UGC7751    &   123511.8+410339 &16.5  & 9  &7.90  &$-$13.89  &$-$3.09    &$-$2.43     &$-$10.58\\
AGC749241  &   124001.7+261919 &18.3  &10 & 5.62  &$-$15.28  &$-$4.79    &$-$3.54     &$-$11.25\\
KDG178     &   124010.0+323931 &17.1  &10 & 7.30  &$-$14.23  &$-$3.51    &$-$2.64     &$-$10.67\\
NGC4656UV  &   124415.7+321700 &13.0  &10 & 5.40  &$-$13.11  &$-$2.66    &$-$1.11     &$-$11.20\\
UGCA298     &  124655.4+263351 &15.3   &9 &11.00  &$-$12.73  &$-$1.66    &$-$2.23     &$-$10.09\\
KKH82      &   131258.7+414712 &16.4  &10 & 8.06  &$-$15.07  &$-$4.27    &$-$3.55     &$-$11.80\\
PGC2229803  &  132753.1+434855 &16.2   &9 & 7.24  &$-$13.90: &$-$3.19:   &$-$2.89     &$-$10.69:\\
LV J1328+4937& 132831.2+493738 &17.3  &10 & 8.40  &$-$13.79  &$-$2.95    &$-$2.54     &$-$10.15\\
AGC238890   &  133230.3+250725 &16.6   &9 & 6.80  &$-$14.60  &$-$3.94      &$-$       &$-$11.24\\
UGC08647    &  133948.0+311725 &16.3   & 8 & 8.60  &$-$13.30  &$-$2.44    &$-$2.43     &$-$10.18\\
LV J1342+4840& 134220.1+484057 &16.3   & 9 & 8.40  &$-$14.21  &$-$3.37      &$-$       &$-$10.97\\
dw1343+58   &  134307.0+581340 &15.7   & 9 & 6.95  &$-$13.84  &$-$3.17    &$-$2.56     &$-$10.82\\
PGC2448110 &   140457.8+534129 &17.3   &10 & 5.38  &$-$12.59  &$-$2.14    &$-$3.05     &$-$8.94\\
M101$-$df2   & 140837.5+541931 &19.8 &$-$2 & 6.87  &$<-$15.26  &$<-$4.60      &$-$       &$<-$11.34\\
UGC09540   &   144852.0+344243 &16.8   &10 & 9.30  &$-$12.99  &$-$2.05    &$-$2.10     &$-$9.57\\
NGC6503$-$d1 & 175248.4+700814 &18.6   &10 & 6.25  &$-$15.33  &$-$4.73    &$-$4.09     &$-$11.17\\
Pegasus III &  222422.6+052512 &18.7 &$-$2 & 0.21  &$<-$15.26  &$<-$7.52      &$-$       &$<-$11.83\\
PAndAS$-$01 &  235712.0+433308 &17.9 &$-$3 & 0.78  &$<-$15.20  &$<-$6.35    &$<-$6.33     &$<-$12.09\\
PAndAS$-$02 &  235755.6+414649 &18.7 &$-$3 & 0.78  &$<-$15.20  &$<-$6.34    &$<-$6.28     &$<-$11.79\\
\hline
\end{tabular}
\end{table}

\section{OBSERVATIONS AND DATA REDUCTION}
Observations of nearby dwarf galaxies were made on the 6-m
telescope of the Special Astrophysical Observatory of the Russian
Academy of Sciences in 2015--2018 using SCORPIO focal
reducer~\cite{afa2005:Kaisin_n}. Galaxy images were registered
with \mbox{2048$\times$2048} CCD providing a field of view of
$6\farcm1\times6\farcm1$ and a resolution of
$0\farcs18$~pix$^{-1}$. H$\alpha$ galaxy images and galaxy images
in the neighboring continuum were obtained using a narrow-band
H$\alpha$ interference filter with a width of
$\Delta\lambda=74$~\AA\ and effective wavelength of
\mbox{$\lambda_{\rm eff}=6555$~\AA}, and two intermediate-band
filters SED\,607 with $\Delta\lambda=167$\,\AA, $\lambda_{\rm
eff}=6063$~\AA\ and SED\,707 with $\Delta\lambda=207$~\AA,
$\lambda_{\rm eff}=7063$~\AA. Typical exposures were 2$\times$600
and 2$\times$300 s in the H$\alpha$ line and continuum,
respectively. Because of the small range of radial velocities of
nearby galaxies we exposured their H$\alpha$ images with the same
filter. The radial velocities of the objects studied lie in the
interval from $-333$~km\,s$^{-1}$ for PAndAS-01 to
$844$~km\,s$^{-1}$ for UGCA\,298.

We reduced observational data in a standard way using MIDAS
package. After subtraction of the electronic bias frame we
performed flat field correction based on with flat field frames
obtained by taking sky images during twilight. We then removed
cosmic-ray hits, aligned the frames, and subtracted the continuum
normalized by the images of 5--15 stars located in the frame. We
calibrated the measured integrated H$\alpha$ fluxes of the
galaxies by images of stellar spectrophorometric standards
acquired on the same night. Internal H$\alpha$-flux measurement
errors were typically of about 15\%. Variations of atmospheric
conditions was the main source of these errors. The resulting
H$\alpha$ fluxes also included the contribution from the [N\,II]
doublet emission lines, which was small in the case of dwarf
galaxies and we ignored it. Only in the case of three late-type
spiral galaxies we corrected the measured fluxes for the
contribution of the [N\,II] doublet in accordance
with~\cite{ken1983:Kaisin_n}.

\section{RESULTS OF OBSERVATIONS}

We present the images of the observed galaxies in the form of an
atlas shown in Fig.~\ref{fig1:Kaisin_n}. The left- and right-side
images of each galaxy represent the sum and difference of the
H$\alpha$ and continuum images. The angular size of the images
is~\mbox{$4\arcmin\times4\arcmin$}, the arrows show the direction
toward North and East. In some frames interference pattern can be
seen, which is due to imperfect account of the division by flat
field. The bright stars and objects with anomalous color of the
wave in the right-hand images  as residual ``stumps''. We
corrected the measured integrated H$\alpha$-flux of the galaxy,
$F$(H$\alpha$) (in the  units of erg~cm$^{-2}$~s$^{-1}$), for
Galactic extinction in accordance with~\cite{sch2011:Kaisin_n}.
The absorption-corrected value, $F_c$(H$\alpha$), served for
estimating  the integrated star-formation rate in the galaxy
\cite{ken1998:Kaisin_n} in the units of ($M_{\odot}$~yr$^{-1}$):
 \begin{equation}
 \log(SFR)=\log F_c({\rm H}\alpha)+2\log D+8.98,
 \end{equation}
where $D$ is the distance to  the galaxy in Mpc. Here we ignore
the contribution from the flux of the [N\,II] emission doublet and
the internal absorption in the Galaxy, because both these effects
are small for dwarf galaxies~\cite{lee2009:Kaisin_n}, which make
up a for the majority of observed objects. The only exceptions
were three spiral galaxies IC\,2233, UGC\,4704  and NGC\,3432,
seen edge-on, for which we adopted internal extinction in the
H$\alpha$-line equal to  0$\fm$73, 0$\fm$40, and 0$\fm$77,
respectively, in accordance with~\cite{ver2001:Kaisin_n}.

The list of general parameters of observed galaxies is presented
in Table~1. Its columns give: (1)\mbox{---}the name of the galaxy;
(2)---J2000.0 equatorial coordinates; (3)---integrated B-band
magnitude; (4)\mbox{---}mor\-pho\-lo\-gi\-cal type according to
UNGC~\cite{kar2013a:Kaisin_n}; (5)---distance in Mpc according
to~\cite{kar2013a:Kaisin_n}; (6)---the logarithm of our measured
H$\alpha$ flux; (7)---the logarithm of the integrated SFR
according to the above formula; (8)---the integrated SFR of the
galaxy
 \begin{equation}
 \log(SFR_{\rm FUV})=\log F_c({\rm FUV})+2\log D -6.78,
  \end{equation}
determined from its  far ultraviolet flux ($\lambda_e=1539$\,\AA,
FWHM\,$=269$\,\AA) measured with GALEX satellite
\cite{gil2007:Kaisin_n} corrected for internal extinction in the
Galaxy; (9)---specific SFR (sSFR) normalized to unit stellar mass
adopted from the UNGC catalog \cite{kar2013a:Kaisin_n}.  Several
H$\alpha$-flux estimates obtained from observations made through
cirrus clouds are marked by colons.

\section{SOME FEATURES OF OBSERVED OBJECTS}
{\em PAndAS-03, PAndAS-05, PAndAS-50,  PAndAS-55, PAndAS-01, and
PAndAS-02.} Global clusters at the distant periphery of our
neighboring M31 galaxy. No signs of star formation have been
detected in these objects wither in H$\alpha$-images or in the
FUV-flux.

{\em Pisces\,A, Pisces\,B.} Two isolated gas-rich dwarf galaxies
discovered by Tollerud et al.
\cite{tol2015:Kaisin_n}. The tip of the red giant branch (TRGB)
distances to these galaxies were measured with the HST
\cite{tol2016:Kaisin_n}.

{\em AGC\,112454.} Gas-rich dwarf galaxy from ALFALFA H\,I-survey
\cite{hay2011:Kaisin_n}, likely satellite of the spiral galaxy
NGC\,628. Emission regions in AGC\,112454 have the form of two
touching rings.

{\em N\,672A, N\,672B.} Two likely dwarf satellites of the
NGC\,672 galaxy found on long-exposure images taken with a small
amateur telescope \cite{kar2015b:Kaisin_n}.

{\em Triangulum-II, Wilman\,1, Pegasus-III.} Dwarf spheroidal
satellites of the Milky Way resolved into stars
\cite{lae2015:Kaisin_n,dej2008:Kaisin_n,kim2015:Kaisin_n}.

{\em AGC\,124056.} Gas-rich low surface brightness dwarf galaxy
from ALFALFA survey. A bright star is located north of this
galaxy.

{\em N\,1156dw\,1, N\,1156dw\,2.} Two likely dwarf satellites of
the NGC\,1156 galaxy~\cite{kar2015b:Kaisin_n}.

{\em GALFA-Dw4.} Irregular dwarf galaxy from the catalog of
compact H\,I-clouds~\cite{sau2012:Kaisin_n}.

{\em MADCASH\,J0742+65.} Ultra-dwarf spheroidal ga\-la\-xy, a
satellite of the spiral galaxy NGC\,2403~\cite{car2016:Kaisin_n}.

{\em IC\,2233.} This edge-on  Sd galaxy has a radial velocity and
H\,I-line width of $V_{\rm LG}=586$~km\,s$^{-1}$ and
$W_{50}=178$~km\,s$^{-1}$, respectively. Based on the
Tully--Fisher~\cite{tul1977:Kaisin_n} relation, its distance is
$12.2$~Mpc, which is somewhat greater than the given limit of the
Local Volume ($11$~Mpc). Our estimate \mbox{$\log(SFR_{{\rm
H}\alpha}) = -0.75$} agrees well with the  estimate $-0.71$,
obtained earlier by Kennicutt~et~al.~\cite{ken2008:Kaisin_n}.

{\em N\,2683dw\,2.} Dwarf spheroidal galaxy discovered on an image
obtained with an amateur telescope~\cite{kar2015b:Kaisin_n}, a
likely satellite of the spiral galaxy NGC\,2683.

{\em UGC\,4704.} Gas-rich Sdm-type galaxy seen almost exactly
edge-on. Its distance estimated by the Tully--Fisher relation is
$D=15.2$ Mpc. With its velocity $V_{\rm LG}=584$~km\,s$^{-1}$, the
galaxy is located in the region of Local Velocity Anomaly
\cite{fab1988:Kaisin_n}. Our SFR estimate, \mbox{$\log(SFR_{{\rm
H}\alpha}) = -1.38$}, agrees well with the $-1.42$ value obtained
by Kennicutt et al.~\cite{ken2008:Kaisin_n}.

{\em JKB\,83.} Intergalactic H\,II-region in the M\,81 group.
James et al. \cite{jam2017:Kaisin_n} report an order-of-magnitude
higher H$\alpha$-flux than our measured value. Our repeated images
of JKB\,83 made with the 6-m telescope confirmed its small
H$\alpha$ flux.

{\em N\,3344dw\,1.} A low surface brightness dwarf galaxy in the
vicinity of the spiral galaxy NGC\,3344.

{\em NGC\,3432=Arp\,206=VV\,11.} An Sdm galaxy with distorted
structure. Kennicut et al. \cite{ken2008:Kaisin_n} report a
$0.10$~dex lower H$\alpha$ flux than our estimate.

{\em J1052+3639.} Dwarf satellite in contact with
NGC\,3432. Shows up conspicuously in the GALEX FUV-band.

{\em HS\,1053+3624 and VV\,747.} These are two likely satellites
of NGC\,3432.

{\em Grapes.} An isolated dwarf galaxy with a radial velocity of
 \mbox{$V_{\rm LG}=408$~km\,s$^{-1}$}
\cite{bur2013:Kaisin_n}. Its distance is estimated from its
$V_{\rm LG}$ adopting a Hubble constant of
\mbox{$H_0=73$~km\,s$^{-1}$\,Mpc$^{-1}$}. Our measured  H$\alpha$
flux of this galaxy is $1.07$~dex higher than reported
in~\cite{bur2013:Kaisin_n}.

{\em MAPS\,1206+31.} Blueish emission galaxy with a small
high-contrast nucleus, which breaks up into several compact nodes
on the H$\alpha$ image.

{\em KK\,135.} An isolated dIr galaxy with a radial velocity of
$V_{\rm LG}=326$~km\,s$^{-1}$, implying a kinematic distance of
$4.46$~Mpc. It is a possible peripheral member of the M\,81 group.
On the H$\alpha$ image KK\,135 has the form of an open ring with
one emission-line condensation.

{\em PGC\,5059199=AGC\,749235.} A small blue gas-rich dwarf galaxy
from the ALFALFA survey~\cite{hay2011:Kaisin_n} with a radial
velocity of  \mbox{$V_{\rm LG}=246$~km\,s$^{-1}$} and the
H\,I-line width of \mbox{$W_{50}=98$~km\,s$^{-1}$}. Its structure
is inconsistent with the kinematic distance of $3.37$~Mpc.
Possibly a member of the peculiar  Coma\,I group at the distance
of $16$~Mpc.

{\em DDO\,133.} An Im galaxy at a distance of $4.88$~Mpc
determined from the TRGB \cite{jac2009:Kaisin_n}. The galaxy
contains more than a dozen compact and ring-shaped
H\,II\mbox{-}regions.

{\em NGC\,4656UV.} A dwarf satellite of NGC\,4656, in contact with
the main galaxy and located northeast from it. The galaxy has low
surface brightness in the optical continuum and H$\alpha$ line,
but shows up contrastively in the
FUV-band~\cite{zas2017:Kaisin_n}.

{\em UGCA\,298=Mrk\,1335.} A blue compact dwarf galaxy at the
boundary of the LV ($11.0$~Mpc) with smooth structure and bright
emission-line nucleus. Judging by its \mbox{$SFR_{{\rm
H}\alpha}/SFR_{\rm FUV}\sim2.7$} ratio,\linebreak Mrk\,1335
undergoes the stage of flare activity.

{\em KKH\,82 and PGC\,2229803, LV\,J1328+4937 and LV\,J1342+4840.}
Likely satellites of the bright M\,51 pair of galaxies.

{\em dw1343+58.} Given its radial velocity, this dwarf galaxy is a
satellite of the giant spiral galaxy M\,101 \mbox{($V_{\rm
LG}=365$~km\,s$^{-1}$)}.

{\em PGC\,2448110.} Compact H\,II region at the periphery of the
Sm galaxy NGC\,5474---a satellite of M\,101. In PanSTARRS survey
images~\cite{cha2016:Kaisin_n} compact nodes can be seen inside
this H\,II region. The \mbox{$SFR_{{\rm H}\alpha}/SFR_{\rm FUV} =
8.1$} ratio is indicative of violent starburst phase in his
object.

{\em M\,101-df2.} A new dwarf spheroidal satellite of the M\,101
galaxy. Its TRGB distance, $6.87$~Mpc \cite{dan2017:Kaisin_n},
confirms the membership of this dwarf in the M\,101 group.

{\em UGC\,09540 = KKR\,13.} An isolated gas-rich  dIr galaxy. Its
distance, $9.30$~Mpc, was estimated by the baryonic Tully--Fisher
relation~\cite{kar2017:Kaisin_n}.

{\em KK\,242 = NGC\,6503-D1.} Low surface brightness dwarf galaxy
found by Karachentseva and Karachentsev~\cite{kar1998:Kaisin_n} in
the vicinity of the spiral galaxy NGC\,6503. Koda et al.
\cite{kod2015:Kaisin_n} independently described it as
\mbox{NGC\,6503-D1} and reported a 12 times overestimated
H$\alpha$ flux. According to~\cite{huc2000:Kaisin_n}, the radial
velocity of KK\,242 is \mbox{$V_{\rm hel}=426\pm6$~km\,s$^{-1}$},
its H\,I flux and the line width are \mbox{$F({\rm
H\,I})=2.03$~Jy\,km$^{-1}$\,s$^{-1}$} and
\mbox{$W_{50}=100$~km\,s$^{-1}$}, respectively. These parameters,
which are unusual for a dwarf galaxy, have to be independently
verified.

\section{RESULTS AND DISCUSSION}
We combined the above star-formation rate estimates in LV galaxies
with similar data for other galaxies located within $11$~Mpc. It
was already pointed out in
Ricciardelli~et~al.~\cite{ric2014:Kaisin_n} and
Karachentsev~et~al.~\cite{kar2018:Kaisin_n} that the SFR in
late-type galaxies depends only slightly on the neighborhood. The
trend of the decrease of  SFR from isolated galaxies to group
members barely exceeds the dispersion of this quantity and is most
manifested among the least massive dwarf galaxies. In other words,
SFR in galaxies is mostly determined by their individual
properties rather than by the tidal effects produced by
neighbours.

\begin{figure} 
\setcaptionmargin{5mm} \onelinecaptionstrue \captionstyle{normal}
\includegraphics[width=0.49\textwidth]{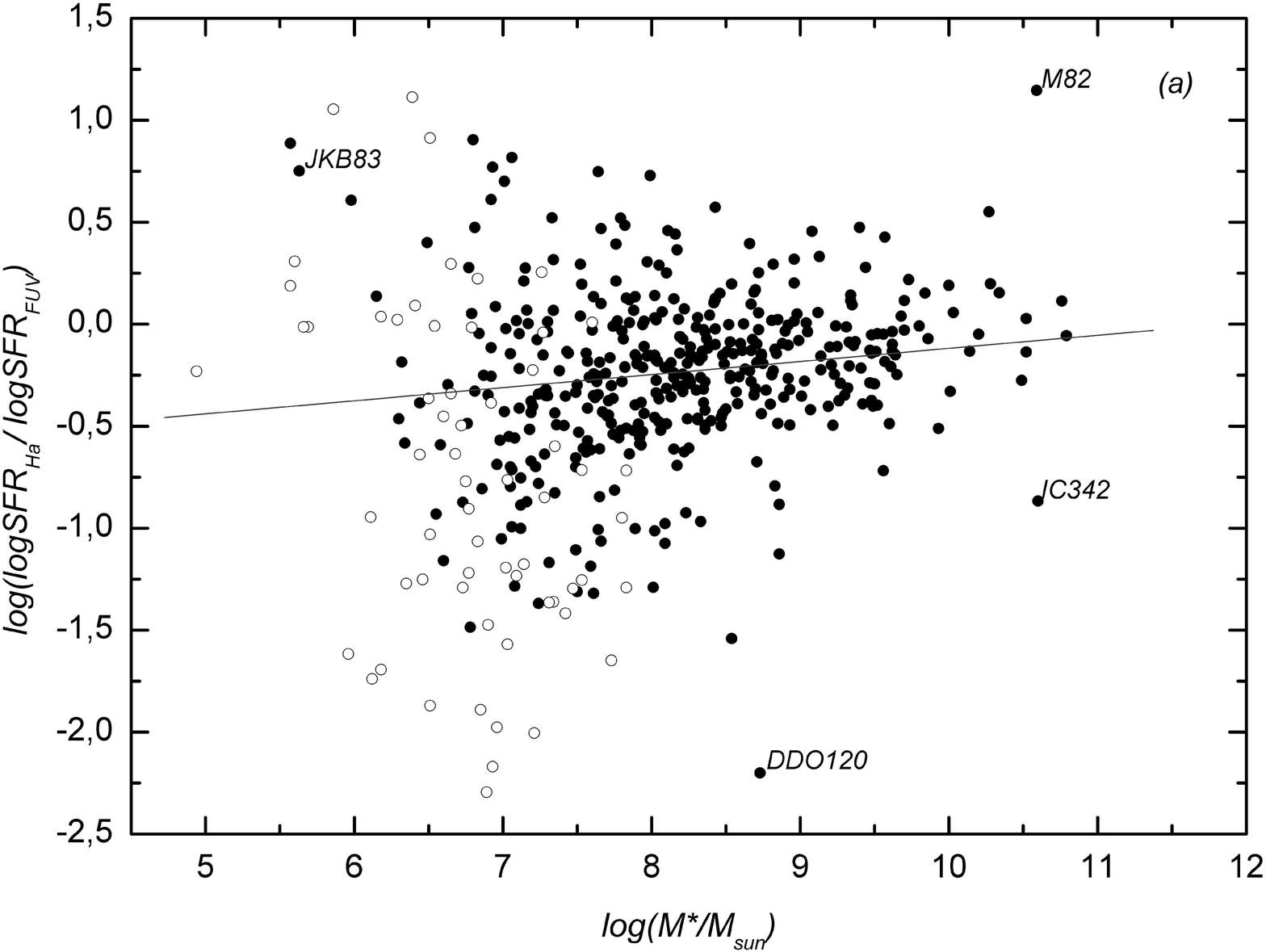}
\includegraphics[width=0.49\textwidth]{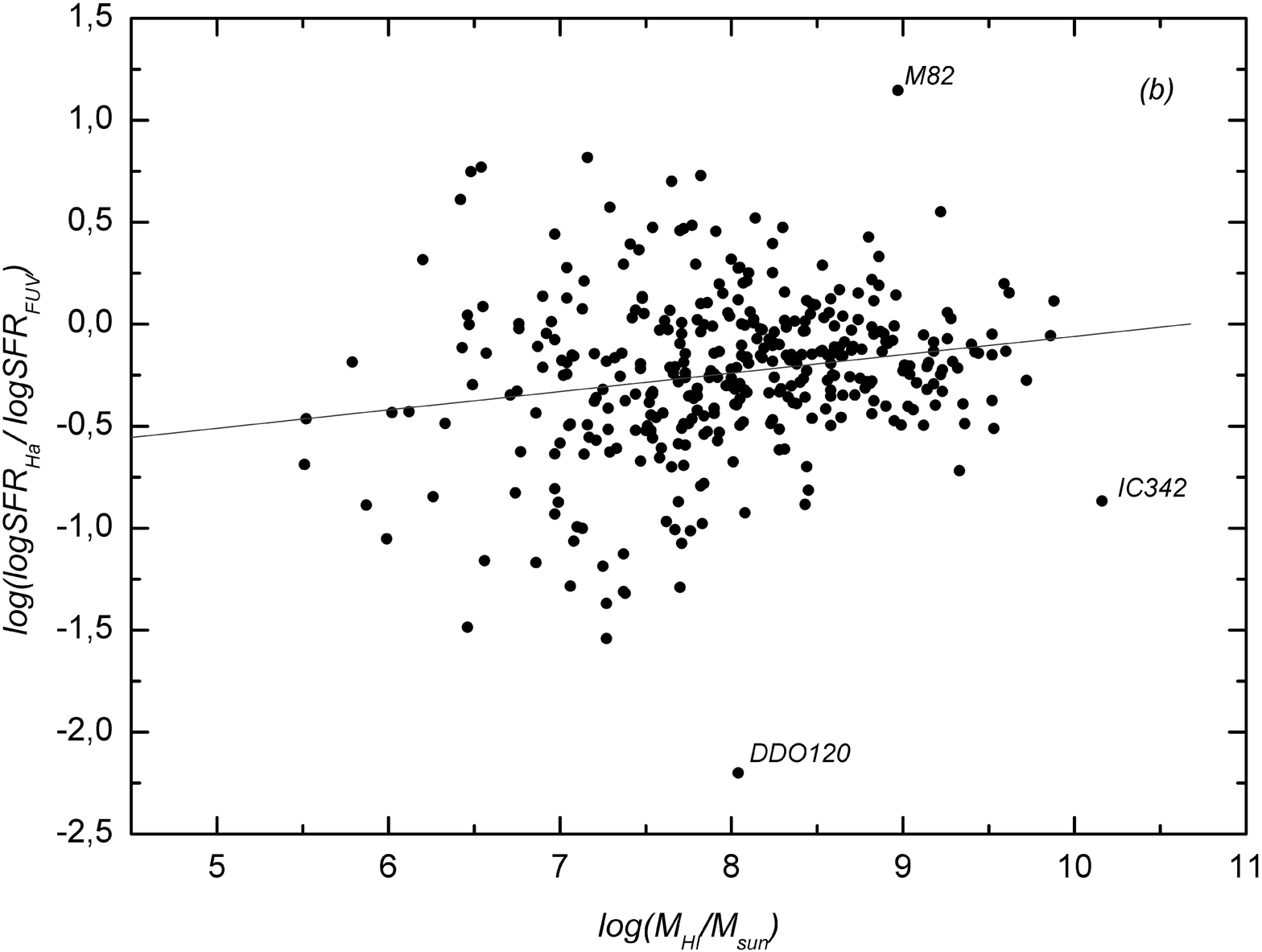}
\caption{The ratio or the SFR determined from H$\alpha$ and FUV
fluxes as a function of stellar mass (a) and hydrogen mass of the
galaxy (b).} \label{fig2:Kaisin_n}
\end{figure}

Fig.~\ref{fig2:Kaisin_n} shows the distribution of the ratio of
SFR estimates determined from H$\alpha$ and FUV fluxes for 506 LV
galaxies. The left- and right-side panels show how this ratio
depends on the stellar and hydrogen masses of galaxies,
respectively. The sample includes only late-type galaxies: T = 10
(Ir), 9 (Im, BCD), 8 (Sdm), 7 (Sd), and 6 (Scd), because
elliptical and lenticular galaxies as well as spiral galaxies with
massive bulges (T$\leq5$) have en essentially different
star-formation history. Galaxies with measured  FUV fluxes but
with only an upper H$\alpha$ flux limit are indicated in
Fig.~\ref{fig2:Kaisin_n} by open circles and we did not took them
into account when computing the statistics of the $SFR_{\rm
H\alpha}/SFR_{\rm FUV}$ ratio. It follows from the data reported
here that the $SFR_{\rm H\alpha}/SFR_{\rm FUV}$ shows a tendency
to decrease toward low-mass galaxies and the dispersion of this
ratio increases significantly from spiral to dwarf galaxies. Such
features of the diagram are due to the burst nature of star
formation in dwarf galaxies, where the amplitude of starbursts on
a \mbox{(10--100) Myr} time scale increases with the decreasing
stellar and hydrogen mass of the
galaxy~\mbox{\cite{ski2005:Kaisin_n,sti2007:Kaisin_n,kar2007:Kaisin_n,mcq2009:Kaisin_n,lee2009:Kaisin_n,kar2013b:Kaisin_n}.}

Particularly striking examples are the M\,82 and Mrk\,475 galaxies
in the starburst phase and the DDO\,120 galaxy in the
post-starburst quiescent phase with their $SFR_{\rm
H\alpha}/SFR_{\rm FUV}$ ratios differing by more than  three
orders of magnitude.

\begin{figure*} 
 \setcaptionmargin{5mm} \onelinecaptionstrue \captionstyle{normal}
\includegraphics[width=0.49\textwidth]{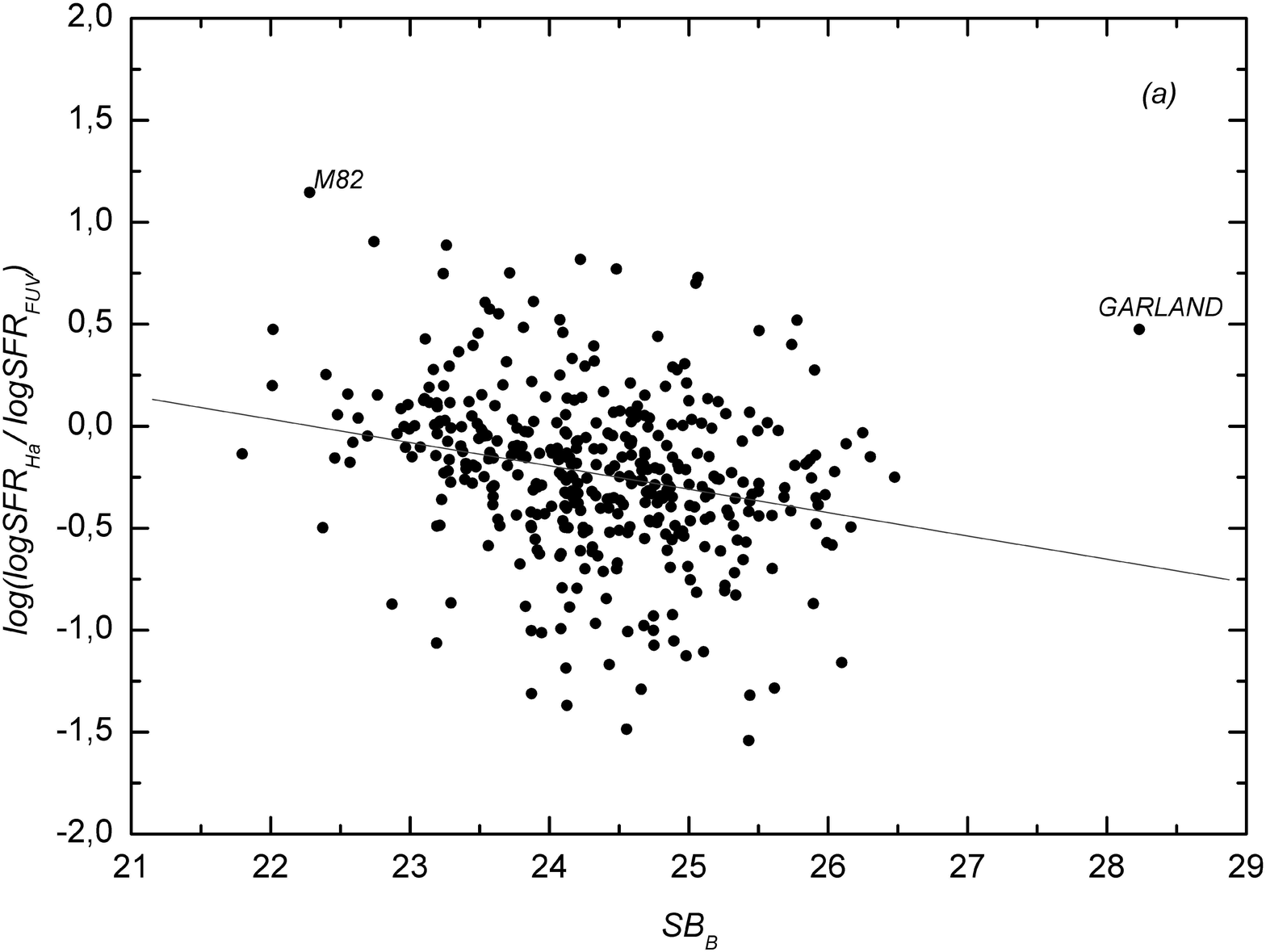}
\includegraphics[width=0.49\textwidth]{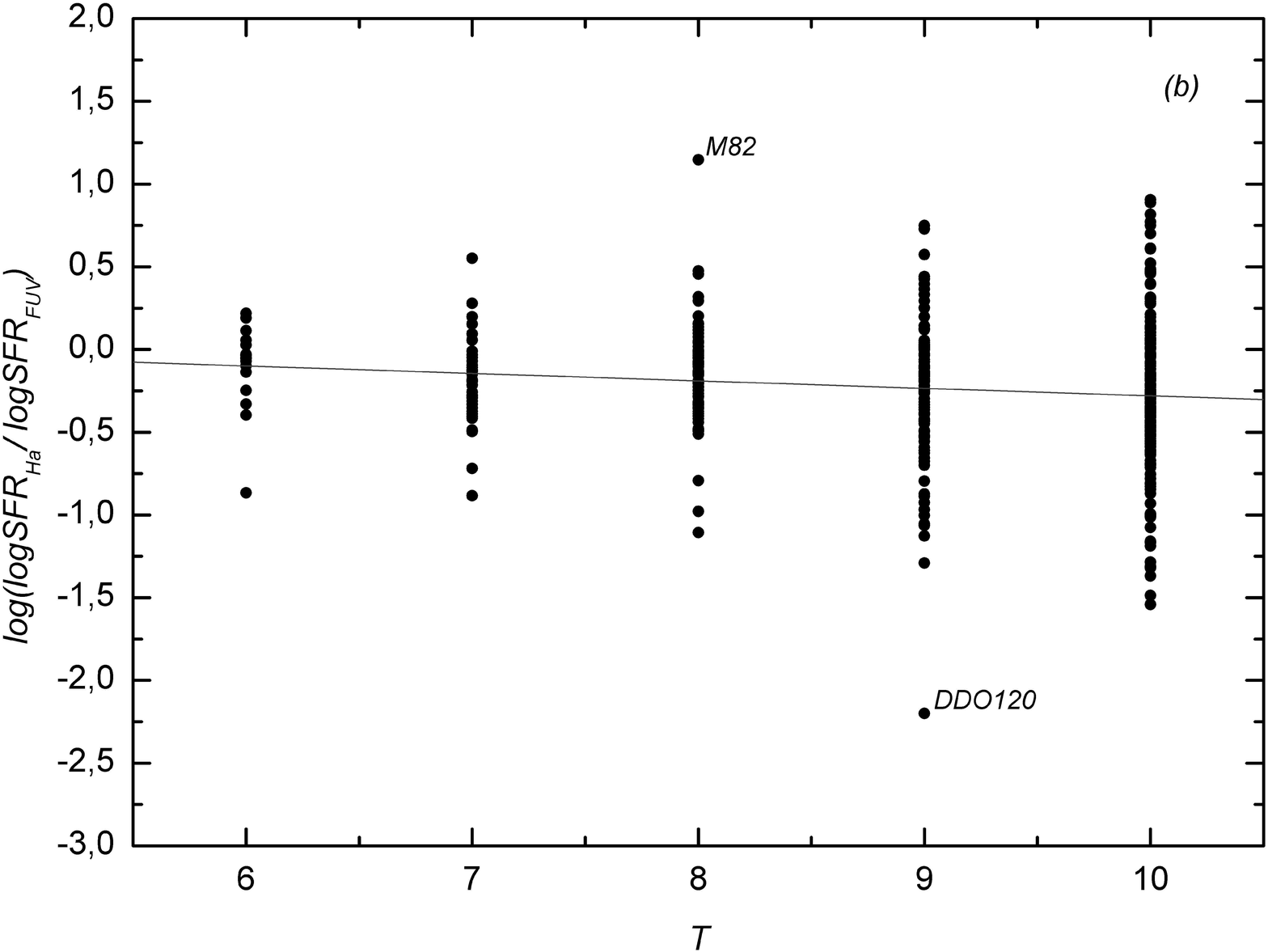}
\caption{Ratio of the SFR estimates determined from  H$\alpha$ and
FUV fluxes as a function of average surface brightness in
magnitudes per square arcsecond ($^{\rm m}/\sq\arcsec$) (a) and
morphological type T (b).} \label{fig3:Kaisin_n}
\end{figure*}

Meurer et al. \cite{meu2009:Kaisin_n} and Audcent-Ross et
al.\cite{aud2018:Kaisin_n} pointed out that the burstlike nature
of star formation in low-mass galaxies shows up in their average
surface brightness. The left-side panel in
Fig.~\ref{fig3:Kaisin_n} shows the dependence of the $SFR_{\rm
H\alpha}/SFR_{\rm FUV}$ ratio on the average B-band surface
brightness of galaxies within the Holmberg radius. As is evident
from the regression line, the average $SFR_{\rm H\alpha}/SFR_{\rm
FUV}$ ratios of BCD galaxies ($SB\sim22^{\rm m}/\sq\arcsec$)
differ from those of low surface brightness dwarfs ($SB\sim26^{\rm
m}/\sq\arcsec$) by about a factor of~$0.5$~dex.

\begin{table}
\setcaptionmargin{0mm} \onelinecaptionsfalse \captionstyle{normal}
\caption{Average integrated parameters and the standard deviations
for galaxies of late morphological types with SFR estimates}
 \medskip
\begin{tabular}{r|r|c|c|c|c|c|c}\hline
  T~  &   N~~  &   $\langle\log M_*\rangle$  &   SD   &   $\langle\log M_{\rm H\,I}\rangle$  &   SD & $\langle\log(SFR_{\rm
H\alpha}/SFR_{\rm FUV})\rangle$&  SD\\
 \hline
  6  &  23  & 9.98$\pm$0.10 & 0.48  &  9.23$\pm$0.08 & 0.39 &  $-$0.11$\pm$0.06 &         0.26\\
  7  &  40  & 9.25$\pm$0.09 & 0.55  &  8.86$\pm$0.07 & 0.45 &  $-$0.19$\pm$0.04 &         0.26\\
  8  &  96  & 8.88$\pm$0.06 & 0.58 &   8.53$\pm$0.05 & 0.49 &  $-$0.15$\pm$0.03 &         0.31\\
  9  & 118  & 8.12$\pm$0.06 & 0.64 &   7.72$\pm$0.06 & 0.69&   $-$0.25$\pm$0.04  &        0.45\\
 10  & 229  & 7.59$\pm$0.04 & 0.64  &  7.56$\pm$0.04 & 0.64 &  $-$0.28$\pm$0.03  &        0.46\\
\hline
\end{tabular}
\end{table}

The right-side panel in Fig.~\ref{fig3:Kaisin_n} reproduces the
distribution of the $SFR_{\rm H\alpha}/SFR_{\rm FUV}$ ratio for
galaxies of different morphological types. Table~2 lists the
average ratio and its standard deviation along with the average
hydrogen mass and its standard deviation for galaxies of different
types. The data in the table indicate that the average $SFR_{\rm
H\alpha}/SFR_{\rm FUV}$ ratio decreases smoothly from spiral to
irregular dwarf galaxies, and the dispersion of this ratio
increases toward dwarf galaxies thereby confirming the idea about
the burstlike nature of star-formation activity in low-mass
objects. Taking into account the LV galaxies for which only the
upper H$\alpha$ flux limit is determined makes this dependence
even steeper and more significant.

The LV contains more than 100 galaxies with measured H$\alpha$
fluxes, which are located outside the coverage zone of the  GALEX
UV survey. The data from Table~2 can be used to obtain an unbiased
estimate of their FUV flux from known  H$\alpha$ flux depending on
the morphology of the galaxy.

The integrated SFR is closely correlated with the luminosity or
mass of the galaxy. Therefore the star-formation process is
usually characterized by the so-called specific star-formation
rate,\linebreak $sSFR=SFR/M_*$, i.e., SFR per unit stellar mass.
Along with it, the ``star-formation efficiency'' parameter,
$SFE=SFR/M_{\rm H\,I}$, is used, which indicates how quickly the
available gas resource in the galaxy will be exhausted.

Given that both the stellar and gas masses of the galaxy vary in
the process of its evolution, it is reasonable to introduce yet
another parameter into the description of evolution without
external influence (the ``closed box'' model)---specific
star-formation rate per unit baryonic mass of the galaxy,
\mbox{$bSFR=SFR/(M_* +1.85 M_{\rm H\,I}$)}. Here the factor~$1.85$
takes into account  the contribution of helium and molecular
hydrogen to the total mass of gas~\cite{fuk2004:Kaisin_n}.

\begin{figure*} 
\setcaptionmargin{5mm} \onelinecaptionsfalse \captionstyle{normal}
\includegraphics[height=0.35\textwidth]{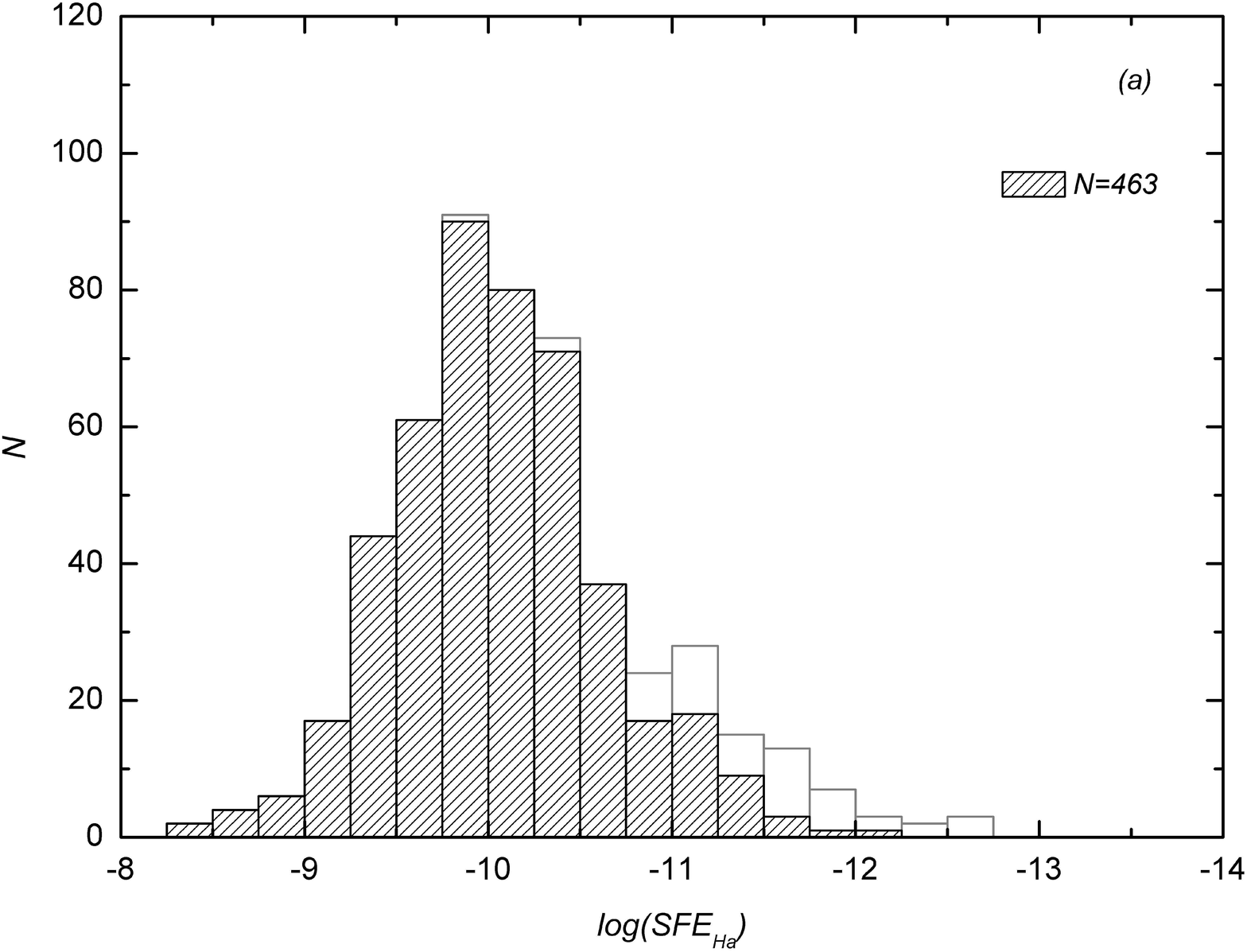}
\includegraphics[height=0.35\textwidth]{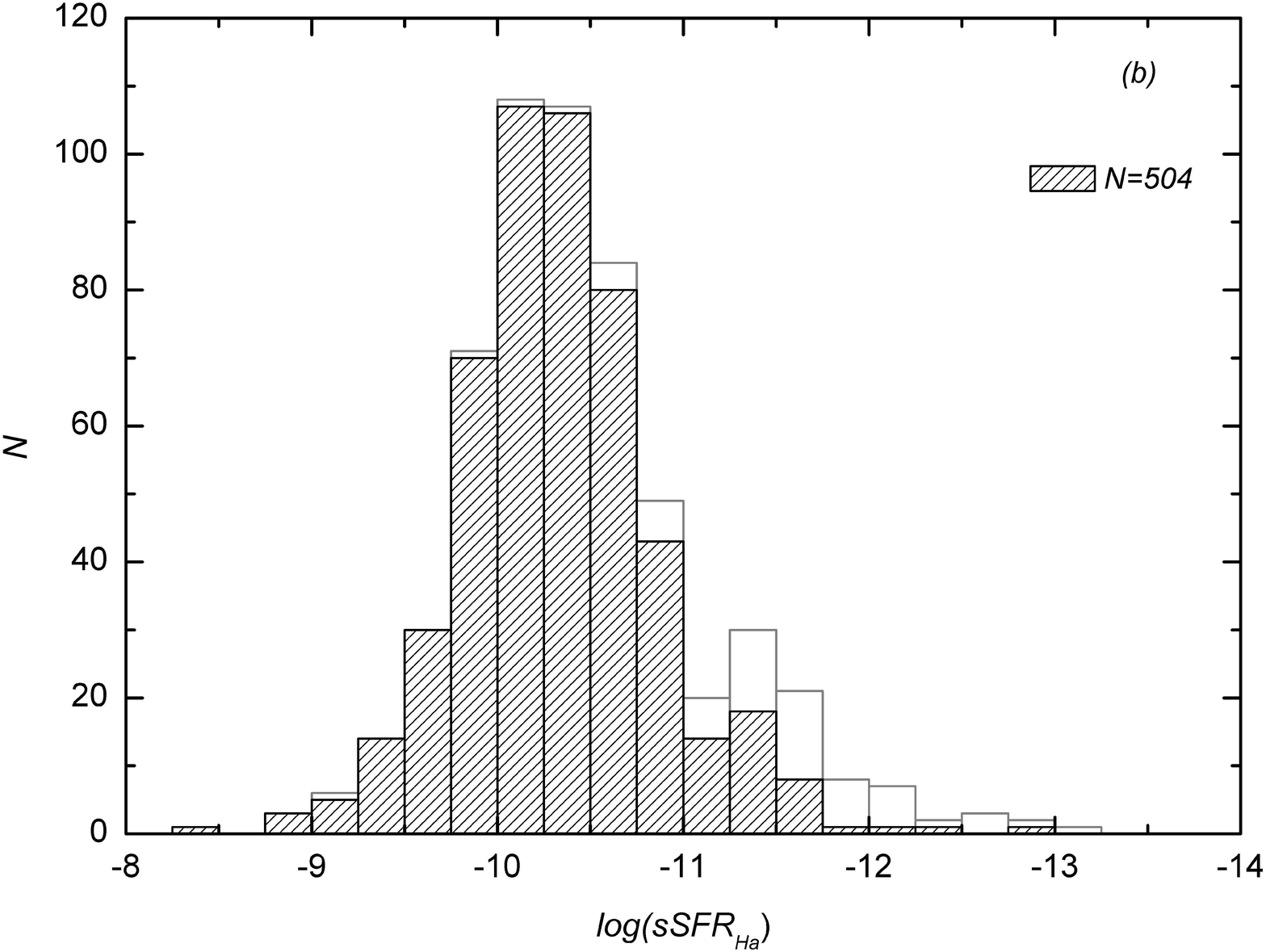}
\includegraphics[height=0.35\textwidth]{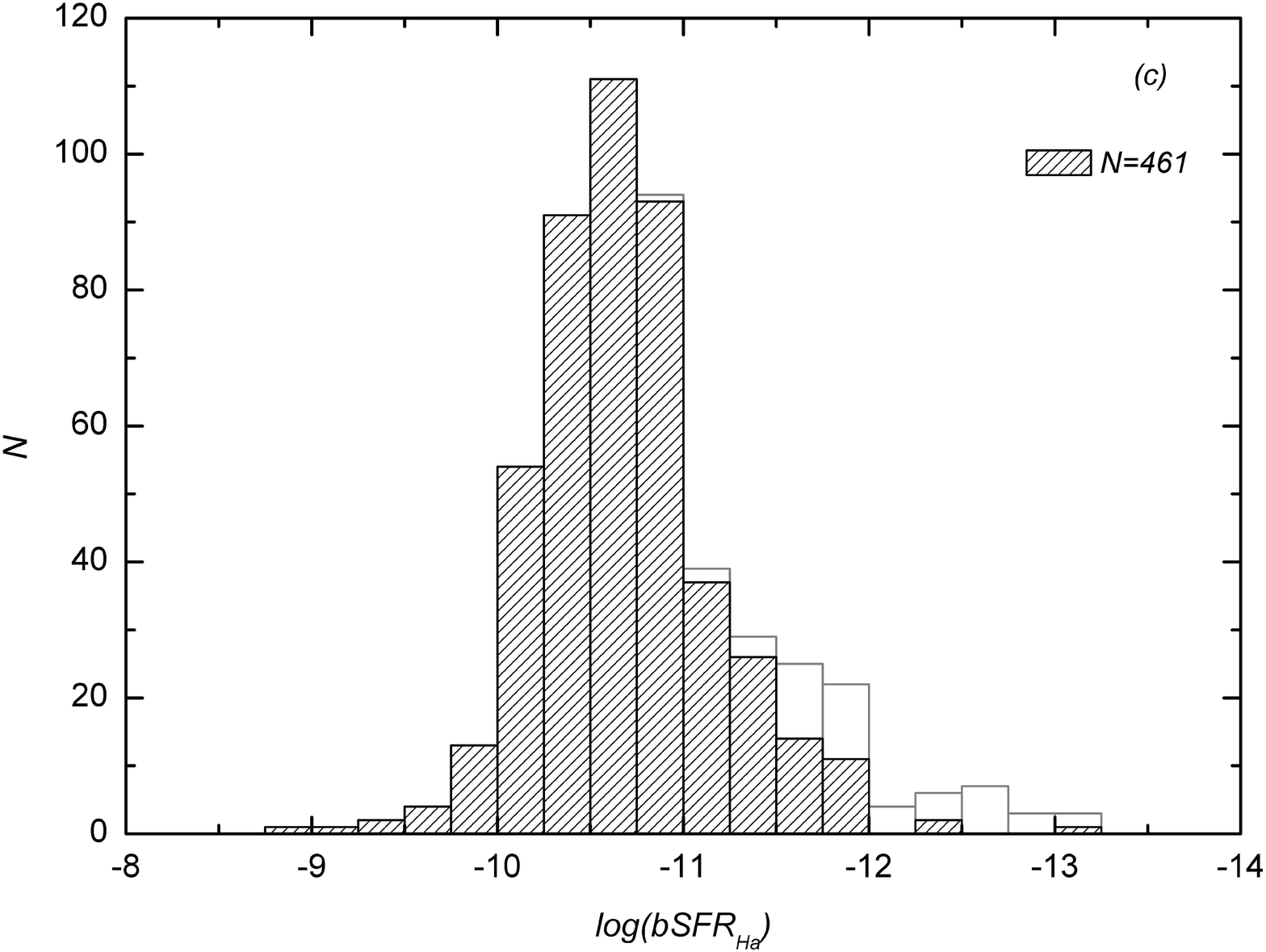}
\includegraphics[height=0.35\textwidth]{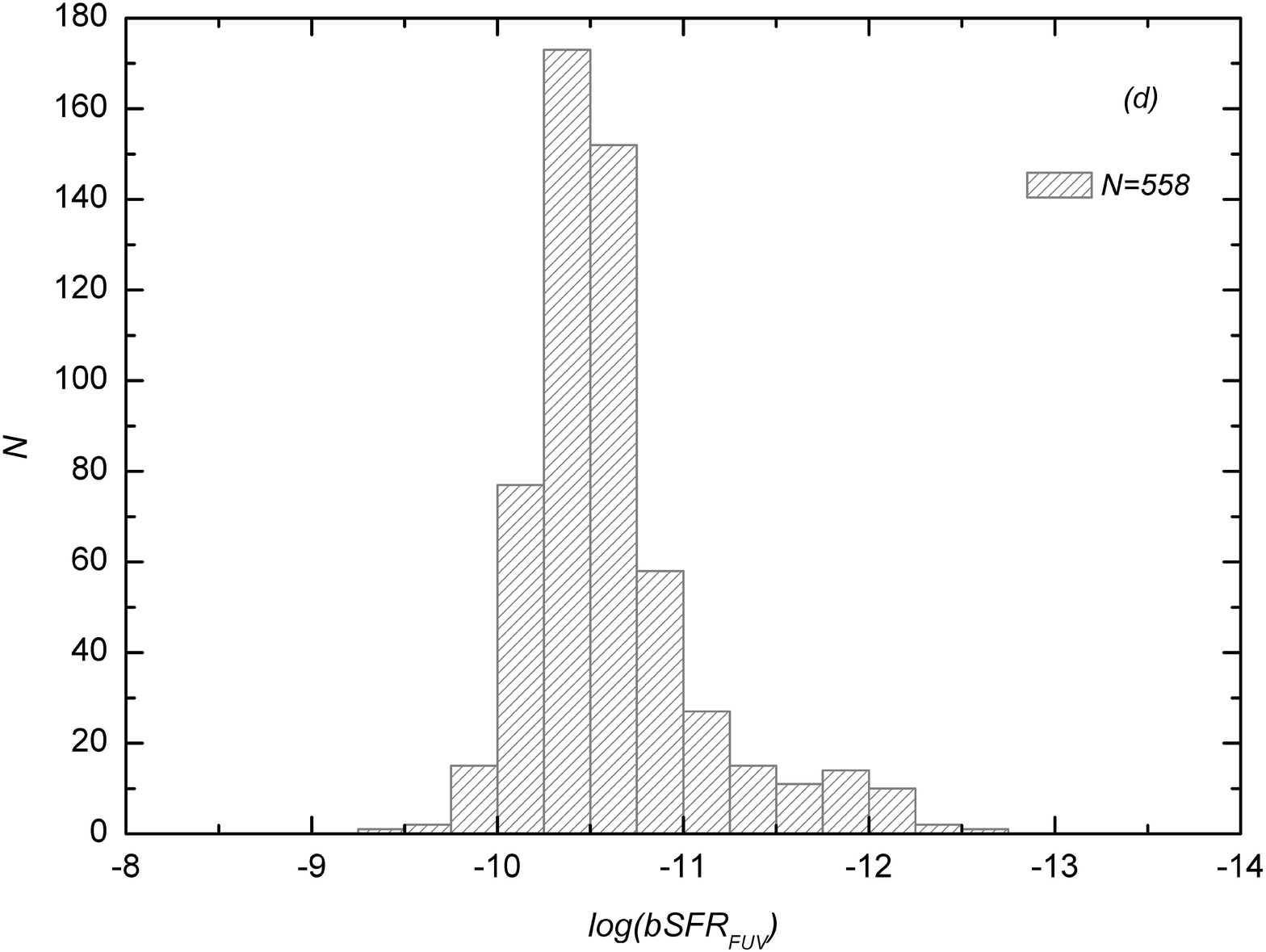}
\caption{Distribution of (a) $SFE$; (b) $sSFR$ determined from
H$\alpha$ flux; (c) $bSFR$ determined from H$\alpha$ flux; (d)
$bSFR$ determined from FUV measurements for LV galaxies. The
horizontal scale is in the units of [yr$^{-1}$] in all plots.}
\label{fig4:Kaisin_n}
\end{figure*}

Fig.~\ref{fig4:Kaisin_n} shows the distribution of the following
parameters of LV galaxies: $sSFR$, $SFE$, and $bSFR$. In each case
SFR is determined from measured H$\alpha$ flux. The distributions
for galaxies with the upper H$\alpha$ flux levels are shown with
unhatched bars. The last plot in the figure corresponds to  $bSFR$
determined from FUV fluxes. An analysis of these histograms leads
us  to the following conclusions.
\begin{list}{}{
\setlength\leftmargin{4mm} \setlength\topsep{2mm}
\setlength\parsep{0mm} \setlength\itemsep{2mm} }
 \item
(a) The distributions of SFRs of late-type galaxies (T\,=\,10--6)
normalized to unit stellar mass, unit hydrogen mass, or unit total
baryonic mass are asymmetric with an excess of galaxies at low
SFR. The inclusion of galaxies with FUV or H$\alpha$ fluxes below
the detection limit (about 10\% of the entire sample) makes the
observed asymmetry even stronger.

\item
(b) The median values of the distributions:\\
$sSFR_{{\rm H}\alpha} = -10.31$ dex, $bSFR_{{\rm H}\alpha} =
-10.64$ dex and $bSFR_{\rm FUV} = -10.52$ dex\\ are close to each
other and are a factor of \mbox{1.5--3} smaller than the expansion
rate of the Universe, \mbox{$H_0 =-10.14$}. The observed SFR in a
typical late-type LV galaxy is quite sufficient for reproducing
most of its observed stellar mass during cosmological time
\mbox{$H_0^{-1}\simeq13.7$ Gyr}. In other words, irregular dwarf
galaxies and disks of spiral galaxies may be evolving in
``glowing'' mode without violent starburst phase at $z\geq2$
typical of early-type galaxies. Note that reserves of gas in type
\mbox{T\,$=$\,10--6} galaxies are sufficient to sustain the
observed SFR in them for almost yet another Hubble time,
\mbox{$0.75H_0^{-1}$}. With the correction due to He and H\,II
content taken into account this quantity increases to
\mbox{$1.4H_0^{-1}$}.

\item (c) The distribution of specific SFR per unit baryonic mass,
$bSFR$, for late-type galaxies has  rather well-defined upper
limit similar to the Eddington limit for stellar luminosity. This
limit is evidently determined by hard feedback: a strong burst of
star formation exhausts local reserves of neutral gas thereby
suppressing further process. The sample of 570 galaxies considered
contains no galaxies whose $bSFR_{\rm FUV}$ would be greater than
$-9.5$~dex. The  $N(\log[bSFR_{{\rm H}\alpha}])$ diagram contains
two compact H\,II-regions: PGC\,2448110  and JKB\,83, and two
compact Markarian galaxies---Mrk\,36 and Mrk\,475---with
$bSFR_{{\rm H}\alpha}$ somewhat greater than the above limit. This
excess, which can be seen in the H$\alpha$ line for about 10~Myr,
diffuses over a time scale of about~100~Myr, which is typical for
the FUV based SFR estimate.
\end{list}

Our H$\alpha$ survey of LV galaxies and the FUV survey of these
galaxies performed with GALEX telescope proved to be rather
sensitive for determination of SFR for $\sim90$\% late-type
galaxies \mbox{T\,$=$\,10--6}.

\section{CONCLUSIONS}

The data presented in this paper lead us to conclude that
star-formation processes in irregular dwarf galaxies and disks of
late-type spiral galaxies have much in common. Most of the
Scd--Sc--Sdm galaxies without apparent manifestations of a bulge
must have never undergone merging acts over about the last 10~Gyr
\cite{kor2010:Kaisin_n}. Such ``virgin'' galaxies are
characterized by regular, sluggish SFR. Dwarf irregular galaxies
have about the same average SFR, however, SFR variations among
them are the higher the smaller is  the baryonic mass of the
dwarf. The SFR and its variations in late-type galaxies are mostly
determined by individual parameters of these galaxies and depend
little on external influences. In the process of the H$\alpha$
survey we found a number of interesting objects where star
formation appears to be caused by external factors, namely, by the
onflow of intergalactic gas \cite{moi2010:Kaisin_n}.

The distribution of specific star-formation rate per unit baryonic
mass,  $bSFR$, for late-type galaxies has a rather well-defined
upper limit similar to the Eddington limit for stellar luminosity,
which must be determined by the presence of hard feedback: a
strong burst of star formation in a galaxy exhausts local reserves
of neutral gas  thereby suppressing further process of the birth
of stars.

 \begin{acknowledgements}
We are grateful to the Reviewer for careful examination of the
paper and the comments that contributed to improving the paper.
\end{acknowledgements}

This work was supported by the Russian Science Foundation (grant
No.~14-02-00965).

{}
\end{document}

%% file: sao_cmd_author.tex
\def\saoname{Special Astrophysical Observatory,  Russian Academy of Sciences,
              Nizhnii Arkhyz, 369167 Russia}

\def\squareforqed{\hbox{\rlap{$\sqcap$}$\sqcup$}}

\def\sq{\ifmmode\squareforqed\else{\unskip\nobreak\hfil
\penalty50\hskip1em\null\nobreak\hfil\squareforqed
\parfillskip=0pt\finalhyphendemerits=0\endgraf}\fi}

\def\arcmin{\hbox{$^\prime$}}

\def\arcsec{\hbox{$^{\prime\prime}$}}

\def\utw{\smash{\rlap{\lower5pt\hbox{$\sim$}}}}

\def\udtw{\smash{\rlap{\lower6pt\hbox{$\approx$}}}}

\def\fm{\hbox{$\,.\!\!^{\rm m}$}}

\def\farcm{\hbox{$\,.\mkern-4mu^\prime$}}

\def\farcs{\hbox{$\,.\!\!^{\prime\prime}$}}

\def\diameter{{\ifmmode\mathchoice
{\ooalign{\hfil\hbox{$\displaystyle/$}\hfil\crcr
{\hbox{$\displaystyle\mathchar"20D$}}}}
{\ooalign{\hfil\hbox{$\textstyle/$}\hfil\crcr
{\hbox{$\textstyle\mathchar"20D$}}}}
{\ooalign{\hfil\hbox{$\scriptstyle/$}\hfil\crcr
{\hbox{$\scriptstyle\mathchar"20D$}}}}
{\ooalign{\hfil\hbox{$\scriptscriptstyle/$}\hfil\crcr
{\hbox{$\scriptscriptstyle\mathchar"20D$}}}}
\else{\ooalign{\hfil/\hfil\crcr\mathhexbox20D}}%
\fi}}

\newcommand{\aap}{Astron. and Astrophys. }

\newcommand{\aaps}{Astron. and Astrophys. Suppl. }

\newcommand{\aj}{Astron.~J. }
\newcommand{\apjs}{Astrophys.~J. Suppl. }

\newcommand{\araa}{Annual Rev. Astron. Astrophys. }

\newcommand{\bsao}{Bull. Spec. Astrophys. Obs. }

\newcommand{\mnras}{Monthly Notices Royal Astron. Soc. }

